\documentclass[12pt]{article}
\usepackage{epsfig,amsmath,amssymb}
\usepackage{cite}
\def\href#1#2{#2}

\def\beq{\begin{equation}}
\def\eeq{\end{equation}}

\begin{document}

\begin{flushright}
{\tt BUHEP-06-02\\INT-PUB 06-04\\hep-ph/0602229}
\end{flushright}

\vskip 2cm

\begin{center}
{\Large \bf Linear Confinement and AdS/QCD} \vskip 1cm

{\bf Andreas~Karch$^1$, Emanuel Katz$^2$, Dam T. Son$^3$, Mikhail A. Stephanov$^4$} \\
\vskip 0.5cm {\it  $^1$ Department of Physics, University of
Washington,  Seattle, WA 98195\\} {\it  $^2$ Department of Physics,
Boston University,  Boston, MA 02215\\} {\it  $^3$ Institute for
Nuclear Theory, University of Washington,  Seattle, WA 98195 \\}
{\it  $^4$ Department of Physics, University of Illinois,  Chicago,
IL 60607 \\}

\medskip

{\tt  E-mail: karch@phys.washington.edu, amikatz@bu.edu, \\son@phys.washington.edu, misha@uic.edu} \\
\medskip

\end{center}

\vskip1cm

\begin{center}
{\bf Abstract}
\end{center}
\medskip
In a theory with linear confinement, such as QCD, the masses
squared $m_{n,S}^2$ of mesons with high spin $S$ or high radial
excitation number $n$ are expected, from semiclassical arguments,
to grow linearly with $S$ and $n$. 
We show that this behavior can be reproduced within a putative 
5-dimensional theory holographically dual to QCD (AdS/QCD).
With the assumption that such a dual theory exists and describes
highly excited mesons as well, we  show that
asymptotically linear $m^2$ spectrum translates into
a strong constraint on the {\em infrared} behavior of that theory.
In the simplest 
model which obeys such a constraint we find $m_{n,S}^2\sim(n+S)$.

\newpage

\section{Introduction}
\setcounter{equation}{0}

Over the recent years it has become clear that
gauge/gravity correspondence \cite{maldacena} can be used to extract
information about four-dimensional strongly coupled gauge theories
by mapping them onto gravitational theories in five dimensions. The
term AdS/QCD is often used to
describe the efforts to apply a
five-dimensional theory on an anti-de Sitter (AdS) gravity
background to learn something about QCD
\cite{wittenqcd,ps,flavor,ps1,csaki,girardello,babington,kruczenski,braga,teramond,amison,pomarol,amimeson,sakai,hirn,hong}.
Although for QCD the exact form of the gravity dual is not yet known,
there are two complementary approaches to the problem. One is to
start from a string theory, choosing the background in such a way as to
reproduce such essential ingredients of QCD as confinement \cite{ps} or
matter in the fundamental representation \cite{flavor}, and study
the resulting QCD-like theories. Another, bottom-up, approach is
to begin with QCD and attempt to determine or constrain the dual
theory properties by matching them to known properties of QCD using
gauge/gravity correspondence.
From a practical point of view, one can model experimental
data surprisingly well~\cite{amison,pomarol,amimeson} by a local effective
theory on a cutoff AdS space (AdS ``slice'').  The ultraviolet (UV) conformal
invariance of QCD (due to asymptotic freedom)
is matched by the conformal isometry of the AdS
background of the dual 5d theory, while confinement, in the simplest
realization, is modeled by a hard wall cutting off AdS space in the
infrared (IR) region, 
as first introduced in \cite{ps1}.
The bottom-up approach is related to an attempt undertaken by A.A.~Migdal
in the 1970's \cite{migdal} to determine the meson spectrum by imposing the
requirement of conformal invariance on QCD two-point correlators
and using the Pad\'e approximation, as well as to the open-moose
models based on an infinite number of hidden local symmetries
\cite{moose}, as discussed in \cite{Erlich:2006hq}.

One criticism that has been brought against this program is that it so
far appeared to be unable to describe correctly either (radially) excited rho
mesons or higher spin mesons~\cite{schreiber,shifman}. The
meson spectrum in AdS/QCD is determined by solving for the eigenmodes
of a 5d gauge field living on the cutoff AdS. With the simplest cutoff
-- the hard IR wall -- the spectrum of squared masses $m_n^2$
is similar to that of a Schr\"odinger equation for a particle in a box,
i.e., for high excitation number, $n\gg1$, $m_n^2$ grow as~$n^2$.

\begin{figure}[ht]
\centerline{\epsfig{file=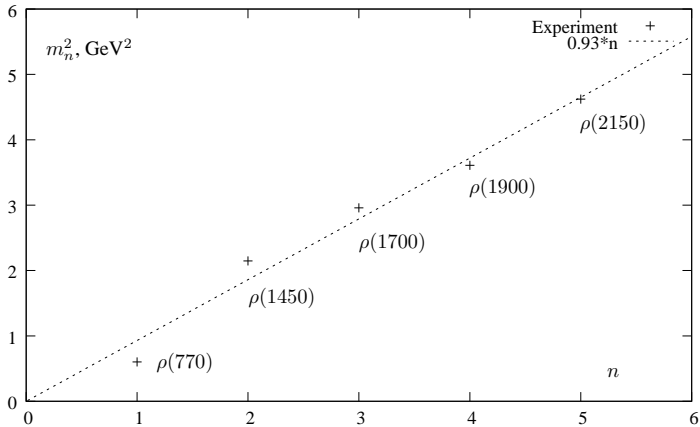,width=0.6\textwidth}}
  \caption[]{The squared masses of the first few $\rho$ resonances
versus their consecutive number $n$ \cite{PDG}. The straight line is
the fit $m_n^2\sim n$.}
  \label{fig:fit}
\end{figure}

On the other hand, data shows growth consistent with $m_n^2 \sim
n$, see, e.g.,
Fig.~\ref{fig:fit} and \cite{collins}.
Furthermore, for large $n$ a heuristic semiclassical
argument in favor of the behavior \mbox{$m_n^2\sim n$} for QCD can be given
\cite{shifman}. The highly excited mesons can be thought of as an
ultrarelativistic quark-antiquark pair executing the semiclassical
motion in a potential growing linearly with the separation (due to the
confining flux tube). With the typical momentum and energy of the
quark motion related to the mass of the meson as $p=E=m_n/2$ and the
energy related to the typical separation $L$ (the size of the meson)
as $E=\sigma L$, where $\sigma$ is the confining string (chromoelectric
flux tube) tension, the
typical size of an excited meson is
\beq L\sim m_n/\sigma \label{size}.\eeq
On the other hand, semiclassical Bohr-Sommerfeld quantization requires
$\int p dx\sim n$, which means
\beq m_n^2\sim \sigma n. \label{mn}\eeq
Such a behavior is also observed in the 1+1 dimensional 't Hooft
model~\cite{hooft} where linear confinement can be demonstrated
analytically.


A similar situation occurs for high spin $S\gg1$.
The well-known argument
based on a picture of a high spin meson as a semiclassically rotating
relativistic open string -- the confining flux tube --
predicts Regge behavior $m_S^2=2\pi\sigma S$.
However, as was shown in \cite{amimeson}, in
that case one finds a very similar problem in the hard wall model:
while one can nicely match the experimental properties of the first
few resonances, the growth of the squared masses with spin $S$ is
$m_S^2 \sim S^2$ as opposed to the expected Regge behavior.

The purpose of this note is to point out that the asymptotic
behavior of the spectrum of highly excited mesons $m_n^2\sim n^2$ is
by no means an intrinsic property of AdS/QCD.  We wish to emphasize
that, contrary to thus far rather common assumption, the spectrum of
the highly excited mesons is not determined by the {\em ultraviolet}
behavior of the AdS/QCD (which is already constrained to be
asymptotically AdS). Rather, it crucially depends on the details of
the {\em infrared} region. That this must be so can be clearly seen
by recalling that the size $L$ of the mesons {\em grows} with their
excitation number \eqref{size}.

Below we shall give an explicit example of the IR wall that gives
the desired
\begin{equation}\label{MnMS}
  m_{n}^2 \sim n \qquad\mbox{and}\qquad  m_S^2 \sim S
\end{equation}
growth of the masses at large $n$ and $S$.
At this point, we can give no explicit example of a background in an
{\it ab initio}
string theory that behaves in the way that we propose.  However, since
the growth~(\ref{MnMS}) is a generic property of any linearly confining
gauge theory, we may formulate our result as an IR constraint on the
holographic dual of any such theory, including QCD. We shall
speculate briefly about how such an IR behavior could arise from
tachyon condensation in string theory.


\section{Background geometry and overview}

The gravitational backgrounds we are interested in can still be
thought of as cutoff AdS spaces, but instead of the hard-wall IR
cutoff we shall look at spacetimes that smoothly cap off. The only
background fields we are considering are the dilaton $\Phi$ and the
metric $g_{MN}$. The mesons are described by 5d fields propagating
on this background with the action given by
\begin{equation}\label{action}
I = \int d^5x\, \sqrt g\, e^{-\Phi}\, {\cal L},
\end{equation}
where $\cal L$ is the Lagrangian density and $g=|\det g_{MN}|$.
We shall begin by considering a generic background parameterized
by two functions $A(z)$ and $\Phi(z)$ such that:
\begin{eqnarray}
&&g_{MN}\,dx^Mdx^N
= e^{2A(z)} (dz^2 + \eta_{\mu\nu}dx^{\mu} dx^{\nu}); \label{metric}
 \\
&&\Phi= \Phi(z);
\end{eqnarray}
where $\eta_{\mu\nu}={\rm diag}(-1,1,1,1)$.
We shall then determine the conditions that the background $A(z)$ and
$\Phi(z)$ should obey to reproduce the Regge-like behavior of the mass
spectrum Eq.~(\ref{MnMS}). By considering the spectrum of radial $\rho$
excitations only we conclude that the linear combination $\Phi-A$ must
behave as $z^2$ at large $z$ to agree with Eq.~(\ref{mn}). In addition,
conformal symmetry in the UV demands that $\Phi-A\sim \log z$ at
small $z$. The simplest solution to both these constraints is
$\Phi-A=z^2+\log z$. It has the advantage that the spectrum of excited
$\rho$ masses can be determined exactly: $m_n^2=4(n+1)$.

In order to determine $A$ and $\Phi$ functions separately we then
consider higher spin mesons. We find that the behavior as in
\eqref{MnMS} requires that the metric function $A$ does not have
any contribution growing as $z^2$ at large $z$.
In the simplest case obeying this constraint,
$A=-\log z$, $\Phi=z^2$, the spectrum can be found exactly:
$m_{n,S}^2=4(n+S)$.

\section{Rho mesons}
\label{sec:rho}


We shall work under the assumption that there exists a local
effective action on this background that is dual to QCD, i.e.,
terms with a higher number of derivatives have to be suppressed.
An incomplete
justification of this assumption together with a discussion of the
associated limitations can be found in Section~\ref{sec:conclusion}.
For the vector and axial meson sector the simplest action one can
write down
containing up to 2 derivatives
is the SU$(N_F)_L \times$ SU$(N_F)_R$ gauge field action with a
bifundamental scalar whose vacuum expectation value is responsible
for both the explicit and the spontaneous chiral symmetry breaking.
This is the picture advertised in \cite{amison, pomarol} and it
directly implements the ideas of \cite{flavor} regarding how flavor
is included in AdS/CFT as gauge fields living on the worldvolume of
flavor branes. The analogy with \cite{flavor} also fixes the
coupling to the background dilaton to be an overall $e^{-\Phi}$ as
expected from a D-brane. The gauge coupling $g_5$ is fixed by
matching the UV asymptotics of current-current two-point function
between bulk and boundary theories \cite{moose,amison}. The action
at quadratic order in the fields and derivatives reads
\beq\label{action-g5}
I = \int d^5x\, e^{-\Phi(z)} \sqrt{g} \left \{
-|DX|^2 + 3 |X|^2 - \frac{1}{4 g_5^2} ( F_L^2 + F_R^2) \right \}
\eeq
with $g_5^2 = 12 \pi^2/N_c$.
The boundary condition on the gauge fields $A_L$ and $A_R$ at
$z=0$ as required by the holographic
correspondence, is given by the value of the sources of the currents
$J_L$ and $J_R$ in 4d theory. The IR boundary condition
now, in the case of the smooth wall extending to $z=\infty$,  is simply
that the action is finite. The ambiguity of the choice of the IR
boundary condition in a theory with hard wall \cite{amison}
is not present in the theory with the smooth wall cutoff.

To determine the spectrum of the $\rho$ mesons we need only the
quadratic part of the action for the vector-like gauge field
$V=A_L+A_R$.  We use the gauge invariance of the action to go to the
axial gauge $V_z=0$ \cite{amison}.  The equation for the 4d-transverse
components $V_\mu^T$ ($\partial^\mu V_\mu^T=0$) has normalizable solutions,
$v_n$, only for discrete values of 4d momentum $q^2$ equal to $m_n^2$:
 \beq \label{rhomode}
\partial_z \left ( e^{-B} \partial_z v_n \right )
+ m_n^2 e^{-B} v_n  = 0,
\eeq
where $B=\Phi(z)-A(z)$. Via the
substitution
\beq v_n=e^{B/2} \psi_n \label{vpsi}
\eeq
this equation can be brought into the form
of a Schr\"odinger equation
\beq
\label{sch} - \psi_n'' + V(z) \psi_n = m_n^2 \psi_n,
\eeq
\beq
V(z) =  \frac{1}{4} (B')^2 - \frac{1}{2} B''.
\eeq

In the particular case of $B=\Phi-A=z^2+\log z$, we have $V=z^2+3/(4z^2)$ and
the Schr\"odinger equation (\ref{sch}) is exactly solvable.
More generally, for the quantum mechanical system%
\footnote{For integer values of $m$ Eq. \eqref{sch-a} can be
viewed as a radial equation for a two-dimensional harmonic oscillator
with orbital momentum $m$.}
\beq \label{sch-a}
-\psi'' + \left[ z^2  + \frac{ m^2-1/4}{z^2}\right] \psi = E \psi
\eeq
the eigenvalues are ($n=0,1,2,\ldots$)
\beq\label{en}
E = 4n + 2 m + 2
\eeq
and the corresponding normalized eigenfunctions are
\beq
\psi_n(z) =  e^{-\frac{z^2}2} z^{m+1/2}
    \sqrt{\frac{2n!}{(m+n)!}}\, L_n^{m}\left(z^2\right).
\eeq
where $L_n^m$ are associated Laguerre polynomials. The $\rho$ meson
mode equation
\eqref{sch} is of this form with $m=1$. We can easily read
off the squared masses of the $\rho$s from this:
\beq m_n^2 =4 (n+1). \label{mn1} \eeq
The scale of the masses is fixed by the coefficient of the $z^2$
term in $B$, which violates explicitly the scale invariance.%
\footnote{In the theory with hard IR wall this role was played by the
position of the wall $z_m$.}
Matching to Eq. \eqref{mn}
we conclude that this coefficient is proportional to
the QCD string tension $\sigma$.
We measure masses in units in which this coefficient is equal to 1.

Undoing the change of variables \eqref{vpsi},
we get the original mode functions
$v_n = e^{z^2/2}\, \sqrt{z}\, \psi_n$, hence
\beq
v_n(z) =   z^{2}
    \sqrt{\frac{2 n!}{(1+n)!}}\, L_n^{1}\left(z^2\right).
\eeq
From the analytic form of the wavefunction we also can read
off the corresponding decay constants \cite{amison}:
\beq
F_{\rho_n}^2 = \frac{1}{g_5^2} \left [ v_n''(0) \right ]^2 =
\frac{8 (n+1)}{g_5^2},
\eeq
whose large $n$ behavior is also in accord with semiclassical QCD
arguments \cite{shifman}.
It is interesting to note that, since $F_{\rho_n}^2/m_n^2=2/g_5^2$ is
$n$-independent, our simplest choice of the background $B=z^2+\log z$
reproduces the {\it ad hoc} resonance model of duality discussed in
\cite{shifman-dual}.

Note that in order to get the correct $m_n^2 \sim n$ behavior for large $n$
it was crucial that the analog Schr\"odinger potential describes
essentially a harmonic oscillator at large $z$. This is easy to see
applying the WKB approximation for large $n$. The distance between successive
levels $m_n^2$ of the Schr\"odinger equation \eqref{sch}
is given by the frequency of the classical oscillation in the potential $V$:
\beq
\frac {dm_n^2}{dn} = \pi \left[\int_{z_1}^{z_2}\!
\frac{dz}{\sqrt{m_n^2-V(z)}} \right]^{-1},
\eeq
where $z_{1,2}$ are the turning points. For large $m_n$, i.e., large
$z_2$, and $z_1\to0$,
the integral is dominated by large $z\sim z_2\sim m_n$. This
matches the expected growth of the
size of the highly excited mesons in QCD  -- $L\sim m_n$
(see Eq.~\eqref{size}).%
\footnote{The proper definition of the meson size should be based on
the meson formfactor, which is determined by cubic terms in the
action. In this paper we only consider quadratic part of the action
and indirectly infer the size of the meson from the extent of its wave
function in the 5-th dimension.}

By choosing a different function $V(z)$ (i.e., a different background
$B(z)$) one can adjust the constant $O(1)$ term in $m_n^2$, but as
long as $V(z)\sim z^2$ for large $z$, the spectrum will remain
equidistant at large $n$.\footnote{One can also notice that the
choice of $B\sim -z^2$ leads to the same asymptotics of $V(z)$ as
the choice $B\sim z^2$ we made. The former, however, is physically
unacceptable because it leads to an $m_n^2=0$ solution to the $\rho$
mode equation \eqref{rhomode}.}

Matching only the spectrum of the $\rho$ mesons we are only able
to constrain the linear combination $B=\Phi-A$ of the dilaton and
the metrics background functions $\Phi$ and $A$. In the next section
we shall see that all the $z^2$ asymptotics must all be in $\Phi$
and none in $A$.%

\section{Higher spin mesons}

In order to create higher spin mesons we need to act with a higher
spin current on the vacuum. Just like for the vector mesons, on the
gravity side, we have to introduce a higher spin field whose
normalizable modes determine meson masses and decay constants.
In the free theory in the far UV the
corresponding current becomes a conserved twist 2 current, so the
higher spin field has to become a massless higher spin field whose
equations of motion are uniquely fixed by gauge and coordinate
invariance \cite{fronsdal}.

In the full theory
the higher spin field will have to acquire a mass in the IR by a
generalization of the Higgs mechanism. As long as the mass of the
higher spin field remains finite in the IR, it is easy to see that it
will not affect the highly excited modes. We will discuss this
phenomenon in detail in the next section in the special case of the
spin $S=1$ axial vectors. For our discussion of higher spin mesons we
will only consider massless higher spin fields in the 5d bulk.

Since we will only be concerned with the spectrum of the higher spin
mesons, we shall not consider the full 5d action, but only its
quadratic (free) part. It is known that simultaneously gauge and
general-coordinate invariant action can be written for a higher spin
field in a space with vanishing Weyl tensor (the part of the Riemann
tensor which does not affect the Ricci tensor)
\cite{deser,fronsdalads,action}. The background we consider
\eqref{metric} is conformally flat and obeys this condition. Thus we
proceed on assumption that such action does exist.

The gauge field of spin $S$ is represented by a tensor
$\phi_{M_1\ldots M_S}$ of rank $S$ totally symmetric over its
indices. As discussed above, we require the action to be invariant
with respect to the gauge transformation with gauge parameter
$\xi_{M_2\ldots M_S}$ itself a symmetric rank $S-1$ tensor: \beq
\delta\phi_{M_1\ldots M_S}=\nabla_{(M_1}\xi_{M_2\ldots M_S)}, \eeq
where $\nabla$ is (general coordinate) covariant derivative and
parentheses denote index symmetrization.%
\footnote{It is known that gauge invariance can be imposed at most
for a restricted class of gauge transformations with traceless
$\xi^N_{\phantom NN\ldots}=0$ gauge parameter. This fact will not
play a role in our discussion.} The quadratic part of the gauge and
coordinate invariant action for this field must have the form \beq
\label{LS} I = \frac12 \int d^5x\,\sqrt g\,e^{-\Phi}\left\{ \nabla_N
\phi_{M_1\ldots M_S}\nabla^N\phi^{M_1\ldots M_S} + M^2(z)
\phi_{M_1\ldots M_S} \phi^{M_1\ldots M_S} + \ldots \right\} \eeq
where the omitted terms are similar to the two written out but with
coordinate indices contracted in alternative ways. The mass
coefficient $M^2(z)$ would be zero for flat space, but must be
nonzero otherwise to cancel terms arising from commutation of
covariant derivatives to ensure gauge invariance of $I$. For the
pure AdS space $M^2(z)$ is a constant $M^2=S^2-S-4$. Using the
method outlined below we can find $M^2(z)$ for a general metric
\eqref{metric}, however, as we shall see, the same method gives the
mode equation directly, bypassing
 $M^2(z)$.

Firstly, similar to the $S=1$ case (Section \ref{sec:rho}), we
utilize the gauge invariance to go over to the axial gauge
$\phi_{z\ldots}=0$. In this gauge, the part of the action involving
the transverse and traceless part of the field $\phi$
($\partial_\mu\phi^\mu_{\phantom\nu\ldots}=0$ and
$\phi^\nu_{\phantom\nu\nu\ldots}=0$) decouples. It is also easy to
see that this part of the action only involves the terms explicitly
written out in \eqref{LS}, while the omitted terms do not
contribute.

Secondly, we use the fact that the axial gauge still allows
residual gauge transformations obeying
$\xi_{z\ldots}=0$ and
\beq \label{residual}
\delta\phi_{z\ldots}=
\nabla_z \xi_{\ldots}  + \nabla_{(.} \xi_{\ldots)
z} =  \xi'_{\ldots} - 2 (S-1) A' \xi_{\ldots}=0.
\eeq
This can be easily integrated to find that the $z$ dependence of the
residual gauge parameter $\xi$ is given by
\begin{equation}
\label{resint} \xi_{\mu_2 \ldots \mu_{S}}(z,x^\mu) = e^{2 (S-1) A(z)} \,
\tilde{\xi}_{\mu_2 \ldots \mu_{S}} (x^{\mu}).
\end{equation}

The  action \eqref{LS} for transverse
traceless modes must be invariant under the gauge
transformations with parameter \eqref{resint}.
As in \cite{amimeson}
this requirement is easiest to implement working in terms of a
rescaled higher spin field $\tilde\phi$ defined by
\begin{equation} \phi_{\ldots} =e^{2 (S-1) A}
\, \tilde{\phi}_{\ldots} \label{tildephi}
\end{equation}
This field simply shifts by a value independent of $z$
 under gauge transformations \eqref{resint},
$\delta\tilde\phi_{\ldots}=\partial_{(.}\tilde\xi_{\ldots)}$.
Therefore
the action written in terms of $\tilde\phi$ should contain only
derivatives of  $\tilde\phi$. Thus if we are to substitute
\eqref{tildephi} into action \eqref{LS} we must find
(discarding appropriate boundary terms)
\beq\label{LStilde}
I = \frac12 \int d^5x\, e^{5A}\, e^{-\Phi}\left\{
 e^{4(S-1)A}\, e^{-2A(1+S)}
\partial_N\tilde\phi_{\mu_1\ldots \mu_S}
\partial_N\tilde\phi_{\mu_1\ldots \mu_S}\right\}.
\eeq
The equation for the modes $\tilde\phi_n$ of the transverse
traceless field $\tilde\phi_{\ldots}$ can be now
easily derived from the action
\eqref{LStilde}:
\begin{equation}
\partial_z \left ( e^{(2S-1)A} e^{-\Phi} \partial_z
\tilde{\phi}_n  \right )+  m_n^2\,e^{(2S-1) A} e^{-\Phi}\,
\tilde{\phi}_n =0
\end{equation}
which has the form \eqref{rhomode} with $B=\Phi-(2S-1)A$.

Converting to
Schr\"odinger form using the procedure from Section \ref{sec:rho}
we see that the only way to have the slope $dm_n^2/dn$
independent of $S$ is to keep all $z^2$ asymptotics in $\Phi$
and none in $A$. For
$A=-\log z$ and $\Phi=z^2$ the Schr\"odinger potential reads
\begin{equation}
V(z) = z^2 + 2(S - 1) + \frac{S^2-1/4}{z^2}.
\end{equation}
This has the same form as the potential in Eq. \eqref{sch-a}.
The eigenvalues corresponding to the
squared masses of the mesons now can easily be read off using
\eqref{en}
\begin{equation}\label{mns}
m_{n,S}^2 =  4(n+S),
\end{equation}
which generalizes our result \eqref{mn1} to higher $S$.

We can also read off the UV conformal dimension of the operators
${\cal O}^{\ldots}$ in QCD dual to the higher spin field
$\phi_{\ldots}$. Since the rescaled field $\tilde\phi$ allows a solution
going to a constant at $z=0$ boundary, its boundary value should be
identified with the source of the operator ${\cal O}^{\ldots}$. The
scaling dimension of the field $\tilde\phi$ is
$[\tilde\phi_{\ldots}]=[\phi_{\ldots}]-2(S-1)=2-S$ (in units in
which $[z]=-1$). Thus $[{\cal
O}^{\ldots}]=4-[\tilde\phi_{\ldots}]=2+S$, i.e., the twist is indeed
equal to 2.

\section{The axial sector}

Similar to the $\rho$ mesons, the axial vector meson $a_1$ masses and decay
constants can be obtained from the modes of the axial gauge field
$A=\frac{1}{2} (A_L - A_R)$ in the bulk. Unlike the vector sector
however the axial field picks up a  $z$-dependent 5d mass via the
Higgs mechanism from the background scalar $X$ that
encodes the chiral symmetry breaking~\cite{amison}.
The axial vector meson mode equation, which follows from
Eq.~\eqref{action-g5}, reads:
\begin{equation}
\label{axialmode}
\partial_z \left ( e^{-\Phi(z)} e^{A(z)} \partial_z a_n \right )
+ [m_n^2 - g_5^2\, e^{2A(z)}\, X(z)^2]
e^{-\Phi(z)} e^{A(z)}
a_n(z)  = 0\, .
\end{equation}
The linearized equation of motion for the field $X$ reads:
\begin{equation}\label{vev}
\partial_z \left ( e^{-\Phi(z)} e^{3 A(z)} \partial_z X(z) \right )
+ 3 e^{-\Phi(z)} e^{5 A(z)} X(z)  = 0\, .
\end{equation}
We are looking for a solution with asymptotic form
\beq \label{XUV}
X(z)\ \stackrel{z\to0}{\to}\
\frac{1}{2} M z + \frac{1}{2}\Sigma z^3.
\eeq
around
$z=0$ (in the UV), where the coefficient $M$ is the UV ($z=0$) boundary
condition given by the
quark mass matrix, while the coefficient $\Sigma$ -- the chiral
condensate -- is determined dynamically by the boundary condition
in the IR.
For large $z$ (in the IR)
on the background $\Phi=z^2$ the equation for $X$ becomes
\begin{equation}
X'' - 2 z X' + \frac3{z^2}X = 0 \qquad(z\gg1).
\end{equation}
Two linearly independent
solutions of this equation have asymptotics $e^{z^2}\to\infty$ and
$\exp\{-(3/4) z^{-2}\}\to 1$. Since the equation is linear,
selecting one of the
solutions in the IR (the $X<\infty$ one, of course) gives
 $\Sigma$ simply proportional to $M$. This is not
what one wants in a theory with spontaneous symmetry breaking
such as QCD. It is clear that one has to consider higher order
terms in the potential $U(X,\ldots)$ for $X$ and all other scalar
condensates. Such a potential
would introduce nonlinearity
in the equation \eqref{vev} for $X$ and consequently in the relation between
$\Sigma$ and $M$. In addition, one expects higher order derivative
terms to become important in determining the precise background
at intermediate values of $z$.

For a generic $U(X,\ldots)$ there will be a solution approaching a constant
as $z\to\infty$. For such a solution, as $z\to\infty$,
the term $X''$ becomes negligible (together with all other higher derivatives)
compared to $-2zX'$, and equation behaves as a 1st order, rather than
a 2nd order equation. That means only one parameter family of
solutions exist in the IR region (that parameter being the value
$X(\infty)$).  Continuing each such  solution into the UV we find
corresponding value for $M$ and $\Sigma$, therefore determining
(parametrically) the function $\Sigma(M)$, which is nonlinear for generic
nonlinear $U(X,\ldots)$.

We conclude that the spectrum in the
axial sector depends sensitively on
 the precise form of the 5d potential $U(X,\ldots)$, as well
as other higher order terms.
However, we expect that
in the IR $X$ will have a solution that goes to a constant as
$z\to\infty$. The constant value that $X$ approaches gives the IR
value of the mass of the axial gauge field.
As can be seen from the axial
mode equation (\ref{axialmode}), the contribution from finite $X$ is
suppressed by a factor $e^{2A}=z^{-2}$. Therefore, the large $z$
asymptotics of the Schr\"odinger equation potential $V(z)$ will be
the same as for the $\rho$ mesons. Hence, the slope of the $a_1$
radial excitation trajectory will be the same as the slope for the
$\rho$-mesons $dm_n^2/dn=4$ as expected from semiclassical arguments
in the Introduction.

As we mentioned in the previous section, the higher spin mesons
find themselves in a similar situation. We know that the corresponding
currents will not be conserved in the interacting theory, so the
dual higher spin field has to pick up a mass in the IR and hence our
analysis based on massless higher spin fields will not be valid.
However, as long as the mass of the higher spin field
remains finite in the IR, at large $z$ and hence at large
$n$ the mass term can be neglected and our result
\beq
m_{n,S}^2 \sim 4(n+S)
\eeq
is reliable for large $n$ and $S$.

\section{Conclusions and discussion}
\label{sec:conclusion}

In this paper we demonstrated that
under the assumption of a local 5d bulk description of QCD
there is a smoothing of the IR wall (asymptotically unique) that
gives the right large $n$ and large $S$ behavior for highly radially
or orbitally excited mesons characteristic of linear confinement
$m_{n,S}^2\sim (n+S)$. We have found that such a spectrum can be
achieved in a nontrivial dilaton background $\Phi$ in \eqref{action}
with the following large $z$ (IR) asymptotics: $\Phi\sim z^2$. We
have also found that the metric background function $A$ in
\eqref{action} cannot have a $z^2$ contribution at large $z$ if the
slope of the radial excitation trajectories $dm_{n,S}^2/dn$ is to be
the same for all $S$.

It is interesting to observe that  in such a dilaton/metric background
the slopes of $n$ and $S$ trajectories automatically coincide:
\begin{equation}\label{mnS}
 \frac{dm_{n,S}^2}{dn} = \frac{dm_{n,S}^2}{dS}\,.
\end{equation}
This matches the expectation from QCD if one considers highly excited
mesons as semiclassically oscillating ($n\gg1$) or rotating ($S\gg1$)
strings -- the confining flux tubes connecting quark and
antiquark.  Indeed, the frequencies of the classical oscillatory and
rotational motions of the relativistic
Nambu-Goto open string at the same energy
coincide \cite{zwiebach}. The poles of the Veneziano amplitude also
correspond to the spectrum obeying \eqref{mnS} \cite{collins}.

We also found the size of the mesons growing linearly with their mass
as required by semiclassical arguments
in QCD Eq.~\eqref{size}. In comparison, with hard IR wall at $z_m$,
excited mesons would have all the same size  $L\sim z_m$.

It is worth pointing out that the semiclassical arguments which
picture highly excited mesons as excited confining flux tubes apply
in the limit of large $N_c$ when the string breaking is suppressed.
In the dual theory, the meson coupling $g_5^2=12\pi^2/N_c$ is small
in this limit, making resonances narrow and allowing us to neglect
loop corrections to the action \eqref{action-g5}.

One interesting phenomenological
aspect of the quadratic dilaton is that the coupling~$g_5^2 e^{\Phi}$
becomes strong in the IR. Modes of excitation number $n$ explore $z$
values of order~$\sqrt{n}$. Thus their self-coupling grows as $e^{z^2}
\sim e^n$. For any given excitation number $n$ the modes will be
weakly coupled for sufficiently large $N_c$: $g_{nnn}\sim e^n/N_c$. 
This then suggests that for a fixed large
$N_c$ the meson resonances will become strongly coupled not at $n
\sim N_c$ but already at $n \sim \log(N_c)$.

Let us close with a detailed discussion of the assumptions we made
and the limitations of this line of thought. Our approach
relies on the proposition that QCD can be described
holographically by a local 5d action, that is, the terms 
with a higher number of derivatives
are suppressed. In this local description all higher spin fields
have to be included as elementary fields.
One should think of this as being the spacetime action corresponding
to string field theory
where all excitations of the string get explicitly incorporated as
spacetime fields. 

Since in QCD there is no parametric separation between the string
tension and the mass gap, the dual gravity will have curvatures of
order the string scale and hence there will be no scale separation
 between the massless and massive string modes; it is hence
very natural to include them all on an equal footing. 
However, it appears unjustified from this point of view to expect
the action of those fields to be dominated by the low derivative
terms, that is, to be local. Without separation of scales, it is not
clear what is suppressing the higher derivative terms. And without
that knowledge, it is not clear how to improve systematically on the
approximations made.

Nevertheless, 
our assumption that such a local action 
is sensible from the string-theory point of view
might be supported by a theorem proposed in \cite{witten}, stating
 that at large $N_c$ the action should always be local\footnote{One
can argue that as long as one is only interested in on-shell
properties described by the quadratic part of the action, like the
masses and decay constants, all higher derivative terms are
essentially equivalent to the standard quadratic terms on a modified
background \cite{inprogress}. However we want to claim that in the
effective action fields of all spins propagate on the same
background geometry, which would not be justified by the arguments
in \cite{inprogress} alone.}.
If this claim is correct, it would be the large $N_c$ that is
suppressing the non-local terms, quite contrary to standard AdS/CFT
intuition. 


A conservative point of view on the 5d Lagrangian would be to consider
it as a phenomenologically driven approach, along the line of
Refs.~\cite{amison,pomarol}, intelligently interpolating between the
low-energy and high-energy limits of QCD.  This approach, while 
being inspired
by the AdS/CFT correspondence, may or may not have any direct
relationship to the latter.

The background we found has an AdS geometry with a quadratic
dilation $\Phi\sim z^2$ turned on. Obviously, one interesting
question is how such a background would arise from a string theory.
That is, what other fields have to be turned on in order to make
this a good background for strings to propagate on?
This would require us to construct the classical string theory on a
spacetime with stringy curvature and presumably RR-fluxes turned on.
This type of background has been sought after now for over a decade
without significant progress. While such a construction would
obviously be desirable, our construction shows that as long as the
assumption of locality holds many of the spacetime properties of
such a theory can be reconstructed.

At least qualitatively we can see how the quadratic dilaton might
arise.
Confinement in the gauge theory is believed to be dual to closed
string tachyon condensation in the bulk, so the simplest hypothesis
is that in addition to the dilaton the only other field that is
turned on is a closed string tachyon of a non-critical super string
theory. It was shown in \cite{barton1} that quite generically it is
possible to engineer a closed string tachyon condensation process in
which the only fields turned on are the dilaton and the tachyon,
while all other fields, in particular the metric, can retain the
background values. So it seems to be completely consistent in this
framework to have an undisturbed background on top of which the
closed string tachyon and the dilaton run. Of course from this point
of view it is far from obvious why the dilaton profile should be
quadratic. So in this respect our finding raises the interesting
prospect that QCD data and/or semiclassical arguments using
confining flux tubes can actually be used to gain new insights into
the fascinating process of closed string tachyon condensation!

Last but not least, one might wonder why none of the known string
theory duals to confining gauge theories exhibit the particular IR
behavior we found. The point here is that up to this date the only
4d examples of the gauge/string theory correspondence that are worked
out have the string theory be well approximated by its low energy
supergravity. On the gauge theory side this is reflected in a separation
of scales --- the string tension is much larger than the mass gap in
the theory. Only the modes with masses above the mass gap but below
the string tension are described by fields propagating on the
gravity background. Note that even the confining examples in that
class do not exhibit meson spectrum characteristic of the linear
confinement.
The squared meson masses in these theories grow as $n^2$ and not as
$n$. Of course, at very high $n$, when the masses are much larger
than the string tension one has to find linear growth in $n$. These
modes however are stringy even in the known examples and will see a
different background than the supergravity modes. Our results
indicate that they will have to see the same exponential wall.

\textbf{Acknowledgments}

D.T.S. thanks the Institute for Advanced Study, where part of this
work was completed, for hospitality. M.A.S. thanks the Institute for
Nuclear Theory, University of Washington for hospitality. The work
of A.K. was supported, in part, by the DOE under Contract No.\
DE-FG02-96-ER40956.  The work of D.T.S. was supported, in part, by
the DOE Grant No.\ DE-FG02-00ER41132 and by an IBM
Einstein Endowed Fellowship from the Institute for Advanced Study.  
The work of M.A.S. was supported, in
part, by DOE Grant No.\ DE-FG02-01ER41195 and the Alfred P.\ Sloan
Foundation.

\end{document}